\documentclass[12pt]{article}

\usepackage{graphicx}
\usepackage{amsmath}
\usepackage{amssymb}
\allowdisplaybreaks

\begin{document}

\baselineskip=17.5pt plus 0.2pt minus 0.1pt

\renewcommand{\theequation}{\arabic{equation}}
\renewcommand{\thefootnote}{\fnsymbol{footnote}}
\makeatletter
\def\CR{\nonumber \\}
\def\pt{\partial}
\def\be{\begin{equation}}
\def\ee{\end{equation}}
\def\bea{\begin{eqnarray}}
\def\eea{\end{eqnarray}}
\def\eq#1{(\ref{#1})}
\def\la{\langle}
\def\ra{\rangle}
\def\hyp{\hbox{-}}

\begin{titlepage}
\title{ 
\hfill\parbox{4cm}
{ \normalsize YITP-05-5 \\{\tt hep-th/0502129}}\\
\vspace{1cm} Effective local geometric quantities in fuzzy spaces \\ from heat kernel expansions}
\author{
Naoki {\sc Sasakura}\thanks{\tt sasakura@yukawa.kyoto-u.ac.jp}
\\[15pt]
{\it Yukawa Institute for Theoretical Physics, Kyoto University,}\\
{\it Kyoto 606-8502, Japan}}
\date{\normalsize February, 2005}
\maketitle
\thispagestyle{empty}

\begin{abstract}
\normalsize
The heat kernel expansion can be used as a tool to obtain the effective geometric quantities in 
fuzzy spaces.  Generalizing the efficient method presented in the previous work on the global quantities, 
it is applied to the effective local geometric quantities in compact fuzzy spaces.
Some simple fuzzy spaces corresponding to singular spaces in continuum theory 
are studied as specific examples.
A fuzzy space with a non-associative algebra is also studied.
\end{abstract}
\end{titlepage}

\section{Introduction}
In classical mechanics the metric of a space-time can be definitely determined 
from trajectories of particles, and the metric tensor is the physical degrees of freedom 
of the general relativity. In quantum theory, however, 
the existence of minimal length in semi-classical quantum gravity and string theory 
\cite{Garay:1994en,Yoneya:2000bt}
and another series of arguments \cite{salecker,karolyhazy,Ng:1993jb,Amelino-Camelia:1994vs,
Sasakura:1999xp} suggest that the metric cannot be definitely determined and 
is not an appropriate tool to describe  a space-time at fundamental level.
This is similar in a fuzzy space \cite{Connes, Madore:aq,Landi:1997sh}, and  
a metric is just what determines the low-frequency dynamics of the fields in it.
Even though such an effective metric is not fundamental, it would be
useful in understanding the gravitational properties of a fuzzy space. 

In continuum theory, the coefficients of the asymptotic expansion of 
the trace of a heat kernel ${\rm Tr}(e^{-tA})$ for $t\rightarrow +0$ are given 
by the integrals of the local geometric 
invariants over a space \cite{Vassilevich:2003xt, Elizalde:1994gf, Gilkey:1995mj, Kirsten:2001wz}. 
Therefore, the heat kernel expansion applied to a fuzzy space may contain the information
on its effective geometric properties. 
The heat kernel expansion for a non-commutative torus with the Groenewold-Moyal star product
was studied in \cite{Vassilevich:2003yz}. 
The author has found a power-law asymptotic 
expansion for $t\rightarrow +0$, and 
the heat trace coefficients in this case are fully defined by the heat trace expansion 
for ordinary but non-abelian operators. 
In the previous paper \cite{Sasakura:2004dq}, the present author studied 
the heat trace for a compact fuzzy space. 
For a compact fuzzy space, it is not appropriate to consider an asymptotic
expansion for $t\rightarrow +0$ because of its finiteness.
It was shown that the effective geometric quantities in a compact fuzzy space are found as 
the coefficients of an approximate power-law expansion of the trace of a heat kernel 
valid for intermediate values of $t$.
An efficient method to obtain these coefficients was presented and 
applied to some known fuzzy spaces to show the validity of the method.

The effective geometric quantities studied in the previous paper \cite{Sasakura:2004dq} are global.
In the present paper, the method will be applied to ${\rm Tr}(h\, e^{-tA})$  with 
the insertion of the operator $h$.
If $h$ has a local support in a fuzzy space, the method
will provide the effective local geometric quantities in a fuzzy space.
I will consider some fuzzy spaces including those 
corresponding to singular spaces in continuum theory. 
Since a fuzzy space smoothens the singularities of a continuum space, 
it would be interesting to see the behavior of the local geometric quantities obtained by the method
near the singularities in continuum theory.

In the following section, I recapitulate the method presented in the previous paper
\cite{Sasakura:2004dq},
and generalize it to include the insertion of $h$ and boundaries.
In Section\,\ref{global}, I study the effective global geometric quantities in fuzzy $S^2/Z_n$,
$S^1$, $S^1/Z_2$ and a fuzzy line segment. In all the cases, the method works well.
In Section\,\ref{local}, the effective local geometric quantities are studied for 
the same fuzzy spaces. The method works well, but some complications appear
in some of them.
The fuzzy $S^1$ obtained through the reduction from  the fuzzy $S^2$ 
turns out to be inappropriate, and the fuzzy $S^1$ defined by a non-associative algebra is used instead.
The local geometric quantities of the fuzzy line segment are hard to be determined because of 
the strong singularities at its boundaries.
The final section is devoted to summary and discussions. 

\section{Coefficient functions}
\label{coefun}
In continuum theory, the asymptotic expansion of the heat kernel has the geometric quantities 
as its coefficients \cite{Vassilevich:2003xt,Elizalde:1994gf, Gilkey:1995mj, Kirsten:2001wz}. For a smooth function $h$ and 
a Laplacian $A=-\Delta$ in a space with no boundaries, the asymptotic expansion for 
$t\rightarrow +0$ is given by\footnote{The usage of the lower index $2j$ for the coefficients 
in the present paper is distinct from the previous one \cite{Sasakura:2004dq,Kirsten:2001wz}, but it
is the one used in the references \cite{Vassilevich:2003xt,Elizalde:1994gf, Gilkey:1995mj}.} 
\be
\label{asymp}
{\rm Tr}(h\, e^{-tA})\simeq \sum_{j=0}^\infty t^{j-\nu/2} a_{2j}(h),
\ee
where $\nu$ is the dimension of the space, and 
\bea
\label{avalues}
a_{0}(h)&=&\frac{1}{(4\pi)^{\nu/2}}\int d^\nu x\, \sqrt{g}\ h,\cr
a_{2}(h)&=&\frac{1}{(4\pi)^{\nu/2}}\int d^\nu x\, \sqrt{g}\  h\,\frac{R}6,\\
a_{4}(h)&=&\frac{1}{(4\pi)^{\nu/2}}\int d^\nu x\, \sqrt{g}\ h\,\left(
\frac{1}{180}R^{abcd}R_{abcd}-\frac{1}{180}R^{ab}R_{ab}+\frac{1}{72}R^2-\frac{1}{30}\nabla_a
\nabla^a R\right),\nonumber
\eea
for $j=0,1,2$. 
Therefore, if a Laplacian is given, one can obtain some geometric quantities
and the dimension through the asymptotic expansion of the heat trace ${\rm Tr}(h\, e^{-tA})$. 

For a compact fuzzy space, 
since the number of independent modes is finite,
${\rm Tr}(h\, e^{-tA})$ is a well-defined function on the entire complex plane 
of $t$. Therefore the behavior of ${\rm Tr}(h\, e^{-tA})$ is not comparable 
with the asymptotic expansion \eq{asymp} of continuum theory. 
However it is generally expected that the continuum description is valid well over 
the scale of fuzziness. 
In the previous paper \cite{Sasakura:2004dq}, the heat trace without the operator insertion
($h=1$) was studied for some known fuzzy spaces.
It was shown that ${\rm Tr}(e^{-tA})$ is comparable with the expansion \eq{asymp} in an 
intermediate range $t_{min}\lesssim t \lesssim t_{max}$, 
and the effective geometric quantities can be obtained through the method explained below.  
The minimum value $t_{min}$ comes from the scale over which the fuzziness can be well neglected 
and the description with an effective metric holds well, 
while the maximum value $t_{max}$ comes from the whole size of a fuzzy space.  

Since the qualitative form of the asymptotic expansion \eq{asymp} in continuum theory does not change with the 
insertion of $h$, the method presented in the previous paper \cite{Sasakura:2004dq}
can be used without any essential modifications. 
Following \cite{Sasakura:2004dq}, let me define a function
\be
\label{defoff}
f_h(t)=t^{\nu/2} {\rm Tr}(h\, e^{-tA}).
\ee 
Then the `coefficient functions' are defined by
\be
\label{coefffun}
a_{h,2j}^N(t)=\sum_{i\geq j}^{N-1} \frac{(-t)^{i-j}}{j!(i-j)!} f_h^{(i)}(t),
\ee
where $f_h^{(i)}(t)$ denotes the $i$-th derivative of $f_h(t)$ with respect to $t$.

In \cite{Sasakura:2004dq}, the global case ($h=1$) is considered and 
it was explicitly checked for some fuzzy spaces that,
with a reasonable choice of $N$, the coefficient functions $a_{h=1,2j}^N(t)$ 
take nearly constant values in a certain range $t_{min}\lesssim t \lesssim t_{max}$,
and that these values can be identified as the effective geometric quantities 
in a fuzzy space corresponding to the coefficients $a_{2j}(h=1)$ 
in \eq{asymp} of continuum theory. 
The intermediate range $t_{min}\lesssim t \lesssim t_{max}$ 
will be denoted by the `stable region' in this paper.       

The coefficient functions $a_{h,2j}^N(t)$ depend not only on the reference scale $t$ but
also on $N$, the number of derivatives considered. 
As was discussed in \cite{Sasakura:2004dq}, this $N$ parameterizes the 
order of the Taylor expansion of $f_h(t)$ about the reference scale $t$, 
and describes the tolerance for the approximation of $f_h(t)$ with a power-law expansion.
To compare the asymptotic expansion \eq{asymp} with the behavior of ${\rm Tr}(h\, e^{-tA})$
of a fuzzy space, a certain amount of tolerance must be allowed. This is because
the spectra of the Laplacian in a fuzzy space do not rigorously agrees with those of a continuum theory.
The tolerance is tighter for larger $N$. Therefore,
the parameter range $t_{min}\lesssim t \lesssim t_{max}$ of the stable region depends on $N$, 
and the stable region becomes smaller for larger $N$. 
In the limit $N\rightarrow \infty$, the stable region disappears.

The above property of the coefficient functions makes it a delicate matter how to 
choose the values of $N$. If $N$ is too small, the approximation of $f_h(t)$
with a power-law approximation will be too bad to provide the accurate
effective geometric quantities, while, if $N$ is too large, 
the stable region cannot be found and no effective geometric quantities can be obtained
as explained above.
Fortunately, the dependence on $N$ is not so large at least for the fuzzy 
spaces studied in the previous paper \cite{Sasakura:2004dq}. In a broad range
of $N$, there exists the stable region and the reasonable values of the effective 
geometric quantities can be found. A mathematically more rigorous criterion for
the choice of the values of $N$ should be formulated in future study, 
but, in the present paper, I will just take one of the values of $N$ with the existence of 
the stable region for each specific case. 
  
The heat kernel expansion in a space with boundaries has additional
contributions from the boundaries, and it is in powers of $\sqrt{t}$ \cite{Vassilevich:2003xt,Elizalde:1994gf, Gilkey:1995mj, Kirsten:2001wz}:
\be
\label{asymphalf}
{\rm Tr}(h\, e^{-tA})\simeq \sum_{j=0}^\infty t^{(j-\nu)/2} a_{j}(h).
\ee
Therefore, the coefficient functions for a fuzzy space with boundaries are  
defined as 
\be
\label{coefffunb}
a_{h,j}^{b,N}(t)=\sum_{i\geq j}^{N-1} \frac{(-t)^{i-j}}{j!(i-j)!} \frac{d^i\, f_h(t^2)}{dt^i},
\ee
where the argument of $f_h(t)$ defined in \eq{defoff} is replaced with $t^2$.
These coefficient functions will be used in the study of fuzzy $S^1/Z_2$. 

\section{Global geometric quantities}
\label{global}
In this section, $h=1$ is taken to study the global geometric quantities.
The coefficient functions \eq{coefffun} are determined from the spectra of 
the Laplacian $-A$ and their degeneracy, since the heat kernel trace is expressed as 
\be
\label{globalheatexp}
{\rm Tr}(e^{-tA})=\sum_{l} d(l) e^{-A(l)\, t},
\ee
where $l$ labels the spectra, and $d(l)$ and $A(l)$ are the degeneracy and the spectra of $A$, respectively.
The present method was already applied to the global geometric properties of 
some regular fuzzy spaces in the previous paper \cite{Sasakura:2004dq}. 
On the other hand, the main motivation of this paper resides in the application to the fuzzy 
spaces corresponding to singular spaces in continuum theory and the behavior of the local geometric
quantities investigated in the next section. I will study fuzzy $S^2/Z_n$, $S^1$, $S^1/Z_2$ and a fuzzy line segment.
In all the examples, the global properties are well obtained through the method, while 
the local properties of some of them show complications.  

\subsection{Fuzzy $S^2/Z_n$}
\label{globals2zn}
In \cite{Martin:2004dm}, fuzzy $S^2/Z_n$ is constructed by taking a subalgebra of 
the fuzzy $S^2$ algebra \cite{Madore:1991bw}. When fuzzy $S^2$ is 
constructed on the spin $L$ representation of the $su(2)$ algebra, 
the fuzzy $S^2$ algebra is the algebra of the $(2L+1)$-by-$(2L+1)$ matrices. 
The algebra elements are classified in terms of the integer $su(2)$ quantum numbers $l,m$ as 
\be
\begin{array}{rcl}
\displaystyle \sum_{i=1}^3 [L_i,[L_i,\hat Y_{l,m}]]&=&l(l+1)\,\hat Y_{l,m}, \\
\displaystyle [ L_3,\hat Y_{l,m} ] &=& m\, \hat Y_{l,m},
\end{array}
\ \ \ (0\leq l \leq 2L,\  -l\leq m \leq l),
\ee
where $L_i\ (i=1,2,3)$ are the $su(2)$ generators.
The $L$ parameterizes the fuzziness of the fuzzy $S^2$, and the continuum
limit is $L\rightarrow\infty$. 
Since the $L_3$ quantum number $m$ is conserved under the 
multiplication, it can be consistently restricted to a multiple of an integer $n$ . 
This subalgebra spanned by 
\be
\hat Y_{l,nk},\  (0\leq l \leq 2 L,\ -l\leq nk \leq l),
\ee
where $k$ takes integers, defines fuzzy $S^2/Z_n$. The Laplacian on  
the fuzzy $S^2$, $-\sum_{i=1}^3 [L_i,[L_i,\cdot]]$, operates consistently on the
subalgebra, and hence can be used as the Laplacian in the fuzzy $S^2/Z_n$.
Thus the spectra and the degeneracy of the Laplacian in the fuzzy $S^2/Z_n$ are given by 
\be
\label{lapdegs2zn}
\begin{array}{rcl}
\displaystyle A(l)&=& \displaystyle l(l+1), \\
\displaystyle d(l)&=& \displaystyle 2 \left[ \frac{l}{n} \right] +1,  
\end{array}
\ (l=0,1,\cdots,2L),
\ee
where $[\cdot]$ denotes the integer part. In the continuum limit $L\rightarrow\infty$,
the fuzzy $S^2/Z_n$ approaches the continuum $S^2/Z_n$, which has the conical singularities
on its two opposite poles.

\begin{figure}
\begin{center}
\includegraphics[scale=.48]{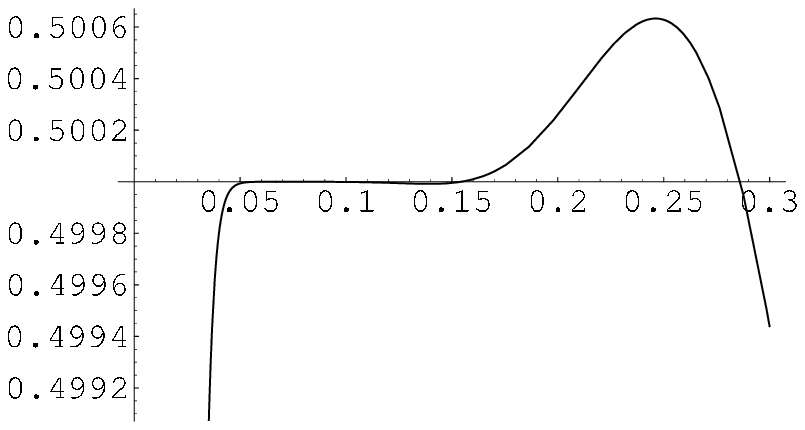}
\includegraphics[scale=.48]{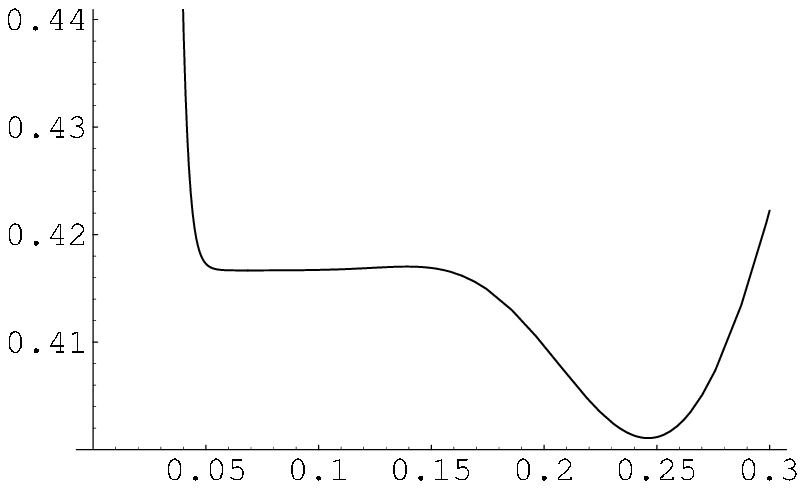}
\includegraphics[scale=.48]{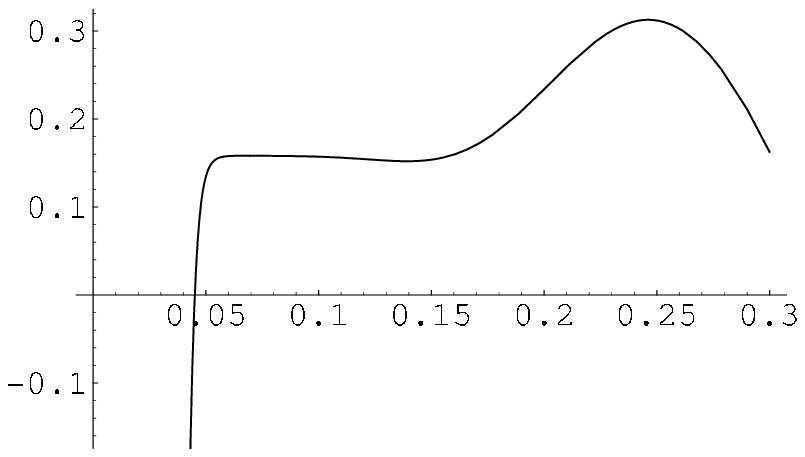}
\caption{The $t$-dependence of the coefficient functions $a^6_{2j}(t)\ (j=0,1,2)$ 
for the fuzzy $S^2/Z_2$. They take almost constant values at $0.05 \lesssim t \lesssim 0.15$. }
\label{fig1}
\end{center}
\end{figure}

As a specific example, let me consider the fuzzy $S^2/Z_2$ with $L=10$. In Fig.\,\ref{fig1}, 
the behavior of the lowest three coefficient functions defined in \eq{coefffun} 
with the choice $h=1$, $\nu=2$ and $N=6$ is shown. 
The stable region is $0.05 \lesssim t \lesssim 0.15$, and their values there may be 
well evaluated at $t=0.08$:
\bea
\label{gs2znavalues}
a^6_0(0.08)&=&0.500000, \cr
a^6_2(0.08)&=&0.416676, \\
a^6_4(0.08)&=&0.158062. \nonumber
\eea

The heat kernel coefficients with $A=-\Delta-\frac14$ on $S^2/Z_n$ in continuum theory are 
explicitly given in \cite{Chang:1992fu} as
\be
\begin{array}{l}
\label{formofck}
\displaystyle {\rm Tr}(e^{(\Delta+1/4)t}) \simeq  \sum_{k=0}^\infty C_k t^{k-1}, \\
\displaystyle 
C_k= \frac{(-1)^{k+1}}{n k!} \left\{ -B_{2k}\left(\frac12\right)+\frac{1}{2k-1}\sum_{m=0}^k
\left(\begin{matrix} 2k \\ 2m \end{matrix}\right) (n^{2m}-1)B_{2k-2m}\left(\frac12 \right)
B_{2m} \right\},
\end{array}
\ee
where $B_k(x)$ is the Bernoulli polynomials defined by
\be
\label{bernoullinum}
\frac{t\, e^{xt}}{e^t-1}=\sum_{k=0}^\infty B_k(x) \frac{t^k}{k!},
\ee 
and the Bernoulli numbers are defined by $B_k=B_k(0)$.
From the expansion \eq{formofck}, the heat kernel coefficients with $A=-\Delta$ are obtained as
\bea
a_0&=&\frac12=0.5, \cr
a_2&=&\frac5{12}\approx 0.416667, \\
a_4&=&\frac{19}{120}\approx  0.158333, \nonumber 
\eea
which are in good agreement with \eq{gs2znavalues}. This supports that the present method
is also applicable to a fuzzy space corresponding to a singular space in continuum theory like an orbifold.

Each coefficient $C_k$ in \eq{formofck} is composed of two distinct kinds of contributions.
The former one is proportional to $1/n$ and can be interpreted as the contribution from the bulk of $S^2/Z_n$.
In fact, it can be obtained from substituting $h=1$, the area of the unit $S^2$, and 
the curvature tensor,
\be
R_{abcd}=g_{ac}g_{bd}-g_{ad}g_{bc},
\ee
into the expression \eq{avalues}, and dividing by $n$.
Then the latter comes from the conical singularities. 
The two distinct contributions can be separately evaluated for $S^2/Z_2$ as 
\bea
\label{locals2z2h0c}
a_{0}({\rm bulk})&=& \frac12, \cr
a_{2}({\rm bulk})&=& \frac1{6}, \\
a_{4}({\rm bulk})&=& \frac1{30}, \nonumber
\eea    
for the bulk, and 
\bea
\label{locals2z2sing}
\Delta a_0&=&0,\cr
\Delta a_2&=&  \frac18, \\
\Delta a_4&=& \frac1{16}, \nonumber
\eea
for one of the singularities.
The separation between the bulk and singularity 
contributions through the present method will be discussed in Section\,\ref{localstwo}.
 
\subsection{Fuzzy $S^1$ and $S^1/Z_2$}
\label{globals1}
As given in \cite{Dolan:2003kq}, fuzzy $S^1$ can be constructed by truncation of 
the fuzzy $S^2$. Consider the limit $\alpha\rightarrow\infty$ of 
the following action of an hermitian scalar field 
in the fuzzy $S^2$ constructed on the spin $L$ representation, 
\be
\label{actionh}
{\rm Tr}\left[\phi [L_3,[L_3,\phi]]+ \alpha \, \phi \left( 2L(2L+1) \phi - 
\sum_{i=1}^3 [L_i,[L_i,\phi]] \right)\right].
\ee
The term $\sum_{i=1}^3 [L_i,[L_i,\phi]]$ has the 
spectra $l(l+1)\ (l=0,1,\cdots,2L)$, and in the limit $\alpha\rightarrow\infty$,
the term multiplied by $\alpha$ in \eq{actionh} works as a constraint, so that
there remain effectively only the modes with $l=2L$ out of all the modes 
$\hat Y_{l,m}$ in the fuzzy $S^2$. 
Therefore the spectra of the fuzzy $S^1$ constructed in this way are given by 
\be
\label{specofasone}
A(m)=m^2,\ (-2L\leq m \leq 2L),
\ee
with no degeneracy. 

\begin{figure}
\begin{center}
\includegraphics[scale=.48]{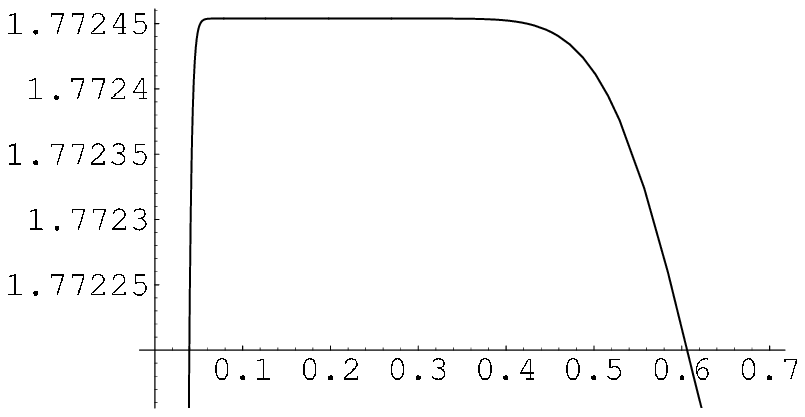}
\includegraphics[scale=.48]{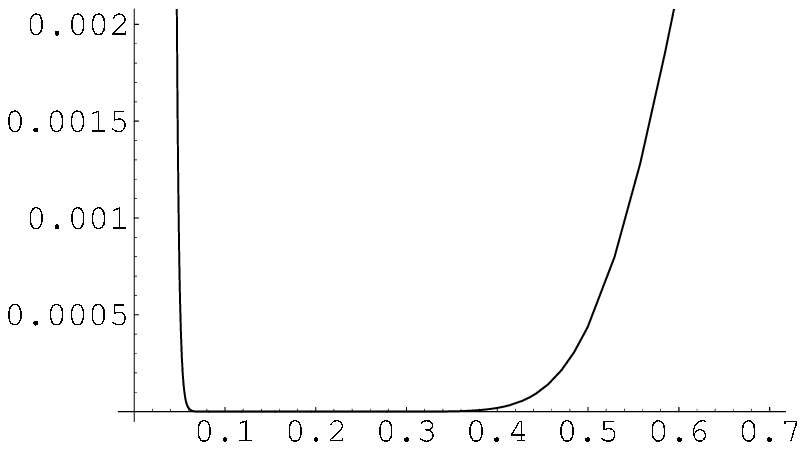}
\includegraphics[scale=.48]{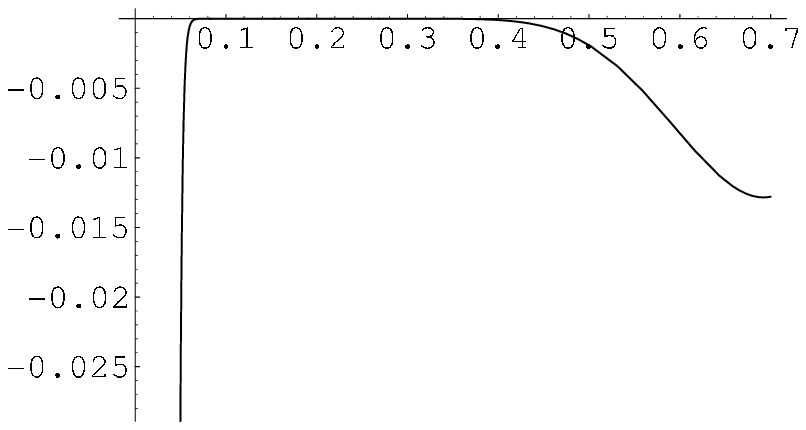}
\caption{The $t$-dependence of the coefficient functions $a^6_{2j}(t)\ (j=0,1,2)$ 
for the fuzzy $S^1$. They take almost constant values at $0.1 \lesssim t \lesssim 0.4$. }
\label{fig2}
\end{center}
\end{figure}

As a specific example, let me consider $L=10$. 
In Fig.\,\ref{fig2}, 
the behavior of the lowest three coefficient functions 
with the choice $h=1$, $\nu=1$ and $N=6$ is shown. 
The stable region is $0.1 \lesssim t \lesssim 0.4$, and their values there may be 
well evaluated at $t=0.2$:
\bea
\label{valuea6}
&&a^6_0(0.2)=1.77245, \cr
&&|a^6_2(0.2)|,\, |a^6_4(0.2)|<10^{-5}.
\eea
These values are consistent with those of the continuum $S^1$ with its total length $2\pi$
obtained from \eq{avalues}.

Next I will consider fuzzy $S^1/Z_2$.
Let me consider the unitary transformation $U$ of parity with 
the property $U^\dagger L_1 U=L_1, U^\dagger L_2 U=L_2, U^\dagger L_3 U=-L_3$. 
It is evidently consistent with the above reduction to the fuzzy $S^1$.
It can be easily shown that the unitary transformation has the eigenvalues $\pm 1$
on the modes of the fuzzy $S^1$, and that the Laplacian has the following spectra 
on the two subspaces of the eigenvalues:
\be
\label{specofas1overz2}
A(m)=m^2,\  \left\{ 
\begin{array}{c}
\displaystyle m=0,1,\cdots, 2L \\
\displaystyle m=1,2,\cdots, 2L 
\end{array}
\right. ,
\ee
with no degeneracy. 

The continuum correspondence of the fuzzy $S^1/Z_2$ is a line segment $[0,\pi]$.
In the continuum limit,
the former case in \eq{specofas1overz2} is the line segment with the Neumann boundary condition 
at the boundaries, 
while the latter the Dirichlet boundary condition.   
The continuum values of the heat kernel coefficients for the Neumann boundary condition
are given by \cite{Vassilevich:2003xt}
\bea
\label{convals1oz2}
a_0&=&\frac{\sqrt{\pi}}2, \CR
a_1&=&\frac12, \\
{\rm Others} &=& 0. \nonumber
\eea
Since there are no bulk contributions to $a_1$ as in \eq{asymp}, 
this non-vanishing value of $a_1$ comes from the boundaries.

\begin{figure}
\begin{center}
\includegraphics[scale=.48]{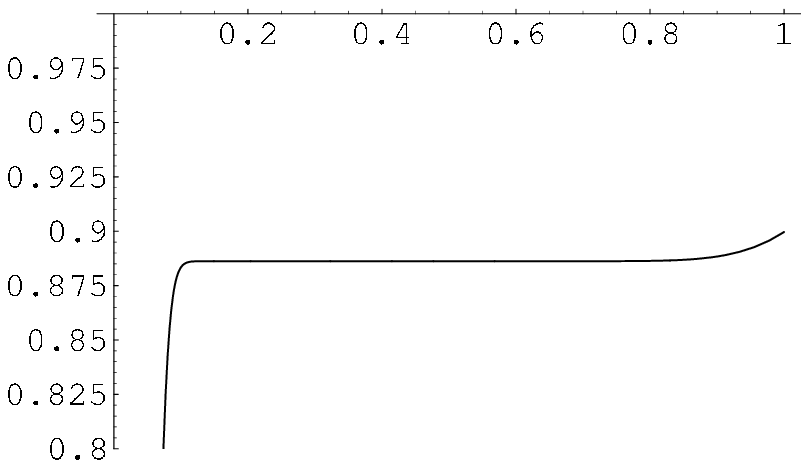}
\includegraphics[scale=.48]{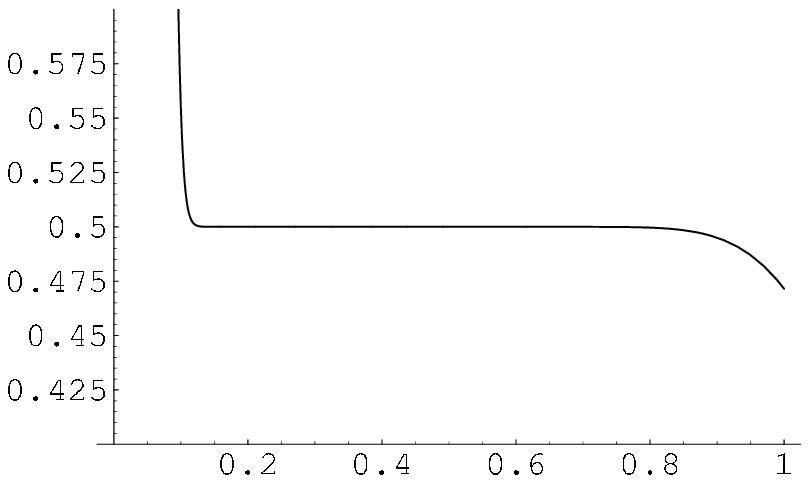}
\includegraphics[scale=.48]{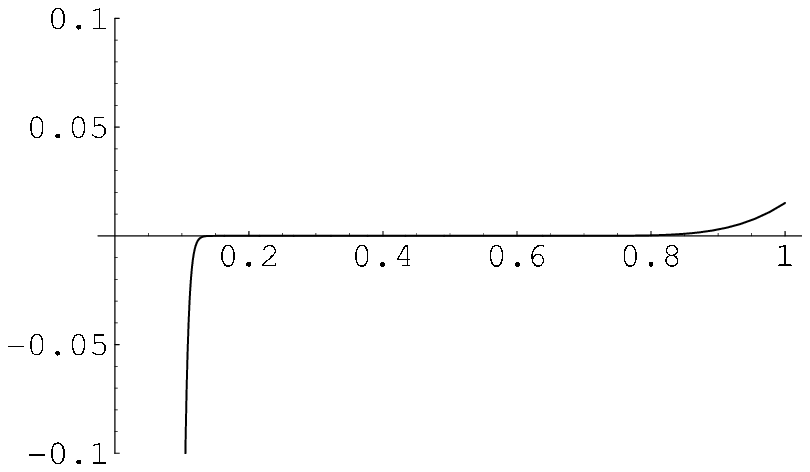}
\caption{The $t$-dependence of the coefficient functions $a^3_{j}(t)\ (j=0,1,2)$ 
for the fuzzy $S^1/Z_2$. They take almost constant values at $0.2 \lesssim t \lesssim 0.7$. }
\label{fig3}
\end{center}
\end{figure}

As a specific example, let me consider the former case $(0\leq m \leq 2L)$ 
in \eq{specofas1overz2} with $L=15$.  
In Fig.\,\ref{fig3}, the behavior of the lowest three coefficient functions 
defined in \eq{coefffunb} with $h=1$, $\nu=1$ and $N=3$ is shown. 
The stable region is $0.2 \lesssim t \lesssim 0.7$, and their values there may be 
well evaluated at $t=0.4$:
\bea
a^{b,3}_0(0.4)&=&0.886227, \cr
a^{b,3}_1(0.4)&=&0.500000,\\
|a^{b,3}_2(0.4)|&<&10^{-14}. \nonumber
\eea
These values are in good agreement with \eq{convals1oz2}.
The separation between the bulk and boundary contributions will be studied
in Section\,\ref{subsections1local}.

\subsection{A fuzzy line segment}
\label{globalseg}
As discussed in \cite{Martin:2004dm, Sasakura:2004yr}, another fuzzy line segment, which is distinct from 
the fuzzy $S^1/Z_2$ in Section\,\ref{globals1}, can be constructed from another truncation of the fuzzy $S^2$.
In this case, the scalar field action in the fuzzy $S^2$ is given by
\be
\label{linesegaction}
{\rm Tr}\left( \sum_{i=1}^3 [L_i,\phi]^2 + \alpha [L_3,\phi]^2 \right).
\ee
In the limit $\alpha \rightarrow \infty$, all the modes with $[L_3,\phi]\neq 0$ are suppressed, 
and the spectra are effectively given by
\be
A(l)=l(l+1),\ (l=0,1,\cdots,2L),
\ee
with no degeneracy. 
The spectra are formally equivalent with the limit $n\rightarrow\infty$ of the fuzzy $S^2/Z_n$. 

\begin{figure}
\begin{center}
\includegraphics[scale=.48]{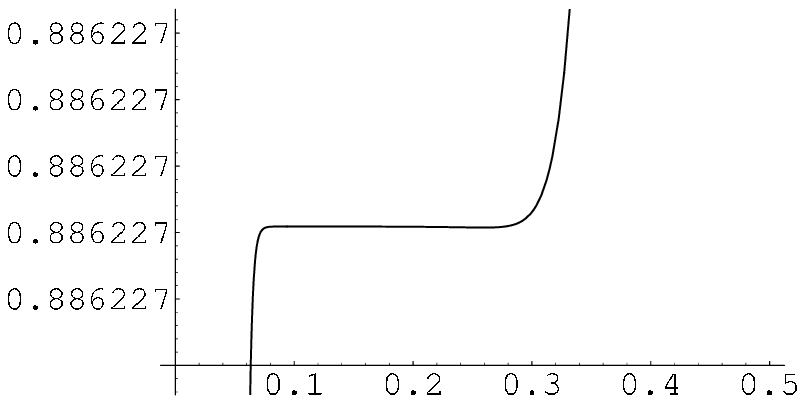}
\includegraphics[scale=.48]{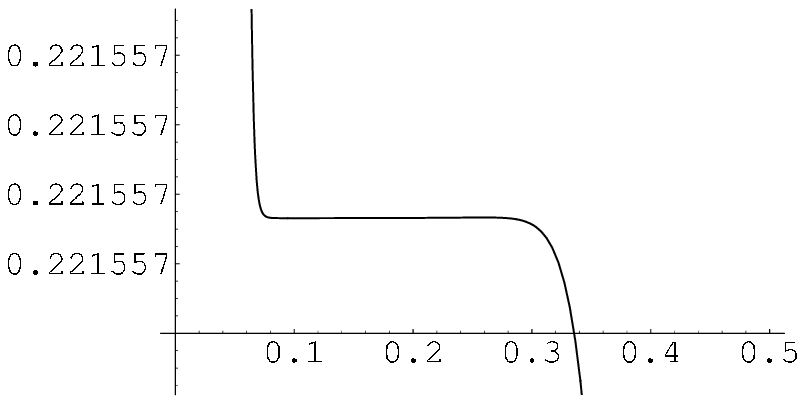}
\includegraphics[scale=.48]{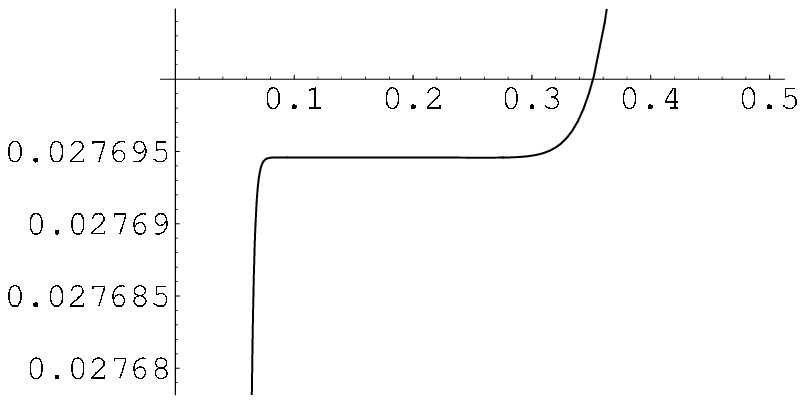}
\caption{The $t$-dependence of the lowest three coefficient functions $a^6_{2j}(t)\ (j=0,1,2)$ 
for the fuzzy line segment with $L=10$. They take almost constant values at $0.1 \lesssim t \lesssim 0.3$. }
\label{fig4}
\end{center}
\end{figure}

The effective global geometric quantities in this fuzzy space were studied in the previous paper 
\cite{Sasakura:2004dq}. The analysis showed that the effective geometric quantities can be 
found from the coefficient functions \eq{coefffun} instead of \eq{coefffunb}, 
which should be used for a space with 
boundaries. In Fig.\,\ref{fig4}, the behavior of the coefficient functions with the choice $h=1$, 
$\nu=1$ and $N=6$ for the fuzzy line segment with $L=10$ is shown. 
There exists the stable region $0.1 \lesssim t \lesssim 0.3$, where the effective
global geometric quantities can be clearly found. 

The above numerical analysis agrees well with the following analytical treatment.
The heat kernel coefficients can be often evaluated 
through the Euler-Maclaurin formula \cite{Vassilevich:2003xt,Elizalde:1994gf, Gilkey:1995mj, Kirsten:2001wz}, 
\be
\label{emformula}
\sum_{l=0}^\infty F(l)=\int_0^\infty d\tau \, F(\tau) +\frac12 F(0)-\sum_{s=1}
\frac{B_{2s}}{(2s)!} F^{(2s-1)}(0) +Rem,
\ee
where $B_{2s}$ are the Bernoulli numbers \eq{bernoullinum}, 
and $Rem$ contains the reminder and the contributions from $F^{(s)}(\infty)$.
Applying the formula to $F(l)=e^{-l(l+1)t}$ and neglecting $Rem$, the asymptotic expansion
for $t\rightarrow 0$ can be obtained as
\be
\label{globalsegasymp}
\sum_{l=0}^\infty e^{-l(l+1)t} \simeq \frac{\sqrt{\pi}}{2}t^{-1/2} +\frac{\sqrt{\pi}}{8}t^{1/2}
+\frac{\sqrt{\pi}}{64}t^{3/2}+\cdots.
\ee
This is in good agreement with the numerical analysis above. 

The global analysis above shows no anomalous behaviors, 
and the global geometric quantities can be safely associated to this fuzzy space.  
However, the local analysis in Section\,\ref{localseg} shows that it is hard to obtain 
the effective local geometric quantities with full conviction. 
The singularities on the boundaries of this fuzzy space
seem to be so strong that the local geometric quantities cannot be obtained through the method,
due to the non-local anomalous property of the boundary contributions.

\section{Local geometric quantities}
\label{local}
In this section, the heat trace with the non-trivial insertion of $h$ will be considered to 
obtain the effective local geometric quantities in the fuzzy spaces studied in Section\,\ref{global}. 

Let me assume that the elements of the algebra defining a fuzzy space are classified by
the eigenvalues of the operator $A$, and denote the elements with the eigenvalue $A(l)$
by $\hat \phi_{l,i}$, where $l$ labels the eigenvalues and $i$ is an additional index labeling the 
independent modes with the same eigenvalue. Let me consider an insertion operator $\hat h$.
It is a natural assumption that $\hat h$ should be an element of the algebra and 
given by a linear combination of $\hat \phi_{l,i}$.  
Then the product of $\hat h$ and $\hat \phi_{l,i}$ can be expressed by a linear combination,
\be
\hat h \, \hat \phi_{l,i}=\sum_{l',i'} c_{l,i;l',i'} \, \hat \phi_{l',i'},
\ee   
with some numerical coefficients $c_{l,i;l',i'}$. The trace operation picks out the coefficients 
in the diagonal, and the heat kernel trace with the insertion can be obtained as 
\be
\label{htgen}
{\rm Tr}\left( h\,e^{-tA}\right)=\sum_{l,i} c_{l,i;l,i}\, e^{-t A(l)}.
\ee 
Thus, not only the eigenvalues but also the algebra defining a fuzzy space 
play significant roles in determining the local geometric properties of the fuzzy space. 
The above procedure would also work for a fuzzy space defined by a non-associative algebra. 
This generalization will be used in Section\,\ref{subsections1local}.

\subsection{Fuzzy $S^2/Z_n$}
\label{localstwo}
The representation space of the spin $L$ representation of the $su(2)$ Lie algebra
is spanned by the vectors $|L,m\rangle \ (m=-L,-L+1,\cdots,L)$, which satisfy
\be
\begin{array}{rcl}
\displaystyle \sum_{i=1}^3 L_i^2\, |L,m\rangle&=&L(L+1)\, |L,m\rangle, \\
\displaystyle L_3\, |L,m\rangle&=&m\, |L,m\rangle.
\end{array}
\ee
It can be easily shown from the $su(2)$ transformation property that
the algebra elements $\hat Y_{l,m}$ in Section \ref{globals2zn} can be explicitly given as
the following operators in the representation space,
\be
\label{repY}
\hat Y_{l,m}=\sum_{m_1,m_2} {C_{l,m}}^{L,m_1;\, L,-m_2} (-1)^{L-m_2} |L,m_1\rangle \langle L,m_2|,
\ee
where ${C_{l,m}}^{L,m_1;\, L,m_2}$ are the Clebsch-Gordan coefficients \cite{Messiah, Varshalovich:1988ye}.
The normalization of these operators is such that
\be
\begin{array}{rcl}
\displaystyle 
{\rm Tr}_L \left( \hat Y_{l_1,m_1}^\dagger \hat Y_{l_2,m_2} \right)&=& \displaystyle \sum_{m_3,m_4}
{C_{l_1,m_1}}^{L,m_3;L,m_4} {C_{l_2,m_2}}^{L,m_3;L,m_4} \\
&=& \displaystyle  \delta_{l_1 l_2} \delta_{m_1m_2},   
\end{array}
\ee
where ${\rm Tr}_L$ denotes the trace over the spin $L$ representation space, and I have
used the reality and the orthogonality of the Clebsch-Gordan coefficients,
\be
\sum_{m_1,m_2} {C_{l,m}}^{l_1,m_1;l_2,m_2} {C_{l',m'}}^{l_1,m_1;l_2,m_2}=\delta_{ll'}\delta_{mm'}.
\ee
Thus the operators $\hat Y_{l,m}$ compose an orthonormal base of the scalar field in the fuzzy $S^2$,
while $\hat Y_{l,kn}$ with integer $k$ compose that of the fuzzy $S^2/Z_n$.

\begin{figure}
\begin{center}
\includegraphics[scale=.59]{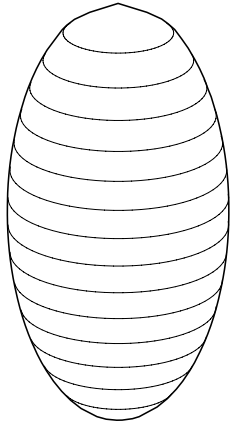}
\caption{Slicing $S^2/Z_n$ into the portions of an equal height. The continuum $S^2/Z_n$ has the conical 
singularities on its two opposite poles.}
\label{fig5}
\end{center}
\end{figure}

Let me define an operator, 
\be
\hat h_{m_1,m_2}=|L,m_1\rangle \langle L,m_2|.
\ee 
When $m_1-m_2$ is a multiple of $n$, this operator is an algebra element of the fuzzy $S^2/Z_n$. 
The identity operator can be expressed as a sum
\be
\label{identityhmm}
1=\sum_{m=-L}^L \hat h_{m,m}.
\ee
Note that $\hat h_{m,m}$ is an algebra element of $S^2/Z_n$.
Since the operator $L_3$ is the fuzzy third coordinate, 
\eq{identityhmm} describes a fuzzy analog of slicing $S^2/Z_n$ into the portions having
an equal height in the direction of the third coordinate, as shown in Fig.\,\ref{fig5}. 

Since the $\hat Y_{l,m}$ compose an orthonormal base, the product $\hat h_{m_1,m_2} \hat Y_{l,m}$
can be expressed as a linear combination of $\hat Y_{l',m'}$:
\be
\hat h_{m_1,m_2} \hat Y_{l,m}=\sum_{l',m'} c_{l',m'} \hat Y_{l',m'}.
\ee
As shown in \eq{htgen}, only the coefficient $c_{l,m}$ is relevant and can be computed from \eq{repY} as 
\be
\label{YhYcoeff}
c_{l,m}={\rm Tr}_L\left( \hat Y_{l,m}^\dagger \hat h_{m_1,m_2} \hat Y_{l,m} \right)=\delta_{m_1 m_2} 
\sum_{m_3} \left({C_{l,m}}^{L,m_1;L,m_3}\right)^2.
\ee
Then the heat trace with the insertion in the fuzzy $S^2/Z_n$ is given by
\be
\label{localtraces2zn}
\begin{array}{rcl}
\displaystyle {\rm Tr}\left( h_{m_1,m_2}\, e^{-t A}\right) &=&
\displaystyle \sum_{l,m} c_{l,m}\, e^{-tA(l)} \\
&=& \displaystyle \delta_{m_1m_2} \sum_{l,m,m_3}  \left({C_{l,m}}^{L,m_1;L,m_3}\right)^2 e^{-t\, l(l+1)}, 
\end{array}
\ee
where I have used \eq{YhYcoeff} and the eigenvalues of the Laplacian \eq{lapdegs2zn}. 
 
When the fuzzy $S^2$ is considered, \eq{localtraces2zn} can be simplified.
The Clebsch-Gordan coefficients has the property,
\be
\sum_{m=-l}^l \sum_{m_3=-L}^L {C_{l,m}}^{L,m_1;L,m_3} {C_{l,m}}^{L,m_2;L,m_3}=\frac{2l+1}{2L+1} \delta_{m_1 m_2}.
\ee
Then \eq{localtraces2zn} is evaluated as 
\be
{\rm Tr}_{S^2}\left( h_{m_1,m_2}\, e^{-t A}\right)=\frac{\delta_{m_1 m_2}}{2L+1}
\sum_{l=0}^{2L} (2l+1)e^{-t\, l(l+1)}.
\ee
Thus the heat kernel with the insertion $h_{m,m}$ is just given by $\frac{1}{2L+1}$
of the global one studied in Section\,\ref{globals2zn}. This shows the uniformity of 
the fuzzy $S^2$, and it is also consistent with the interpretation 
of \eq{identityhmm} as slicing $S^2$ into the portions of an equal height, because each
portion has the same area with the uniform curvature.

\begin{figure}
\begin{center}
\includegraphics[scale=.48]{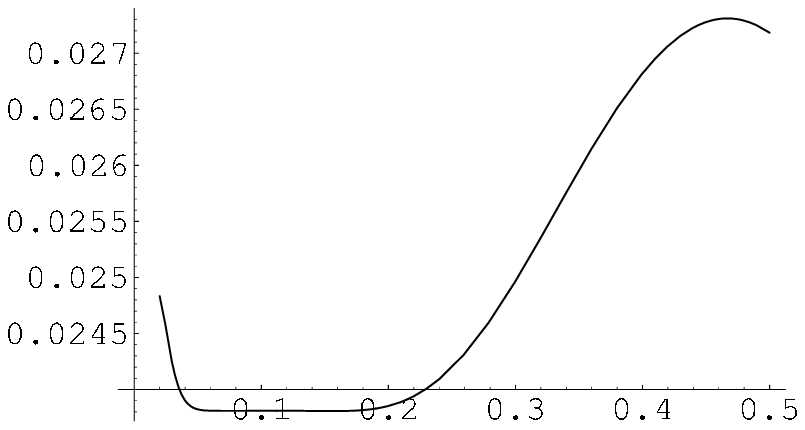}
\includegraphics[scale=.48]{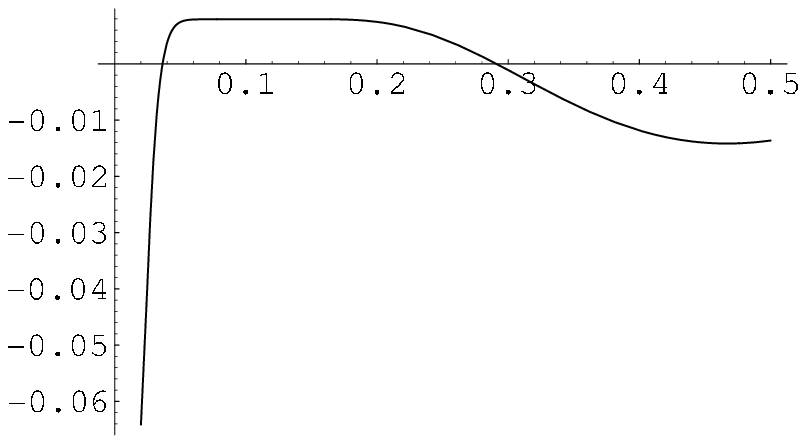}
\includegraphics[scale=.48]{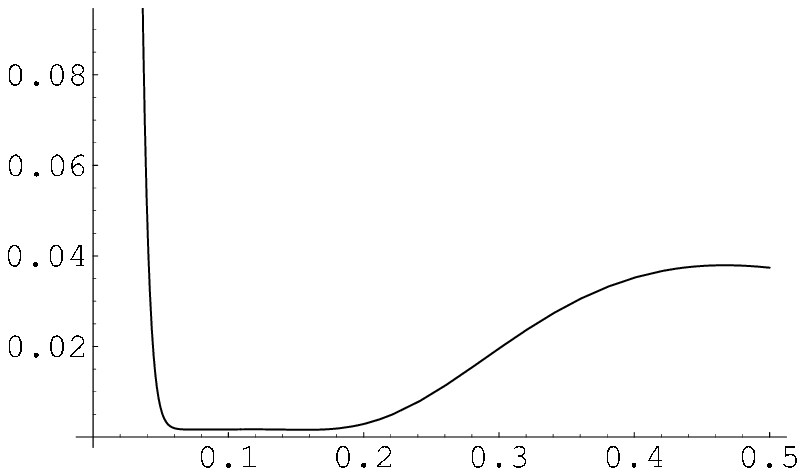}
\caption{The $t$-dependence of the lowest three coefficient functions $a^3_{h_{0,0},2j}(t)\ (j=0,1,2)$ 
for the fuzzy $S^2/Z_2$. The stable region is $0.06 \lesssim t \lesssim 0.2$.}
\label{fig55}
\end{center}
\end{figure}

\begin{figure}
\begin{center}
\includegraphics[scale=.48]{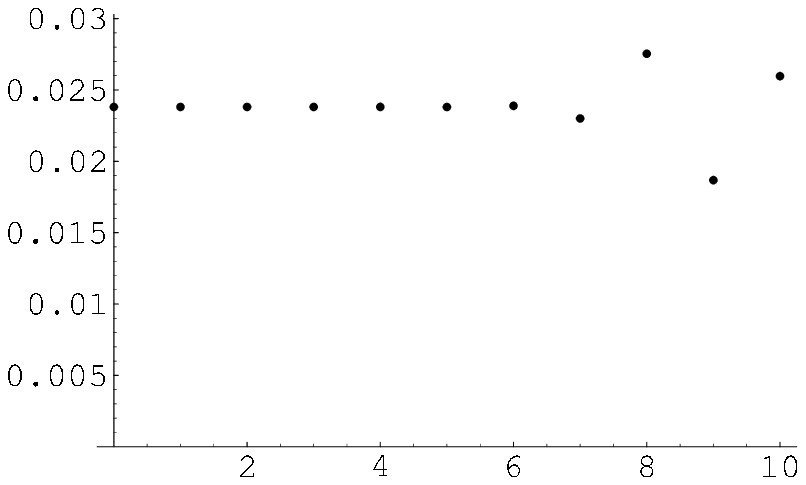}
\includegraphics[scale=.48]{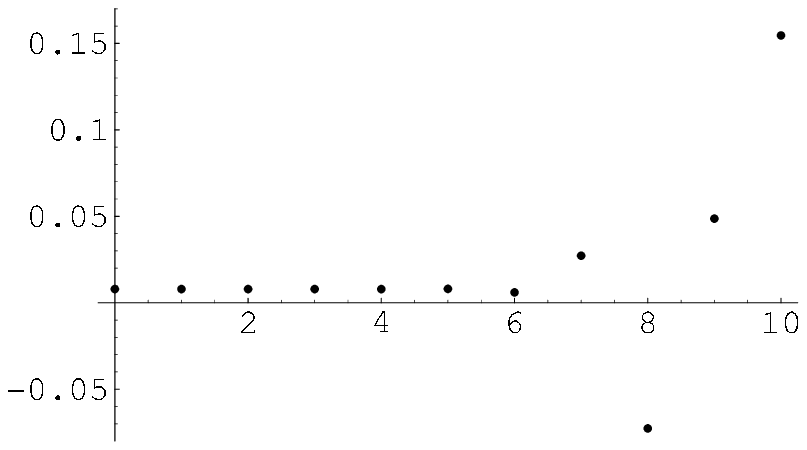}
\includegraphics[scale=.48]{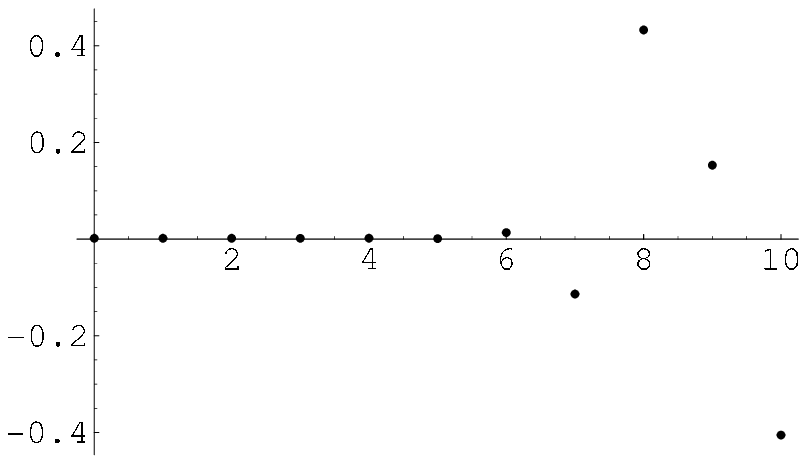}
\caption{The $m$-dependence of $a_{h_{m,m},2j}^{3}(0.08)\ (j=0,1,2)$.}
\label{fig6}
\end{center}
\end{figure}

When the $S^2/Z_n$ is considered, 
the summation of $m$ in \eq{localtraces2zn} is restricted to multiples of $n$:
\be
\label{localtraces2znexp}
{\rm Tr}_{S^2/Z_n}\left( h_{m_1,m_2}\, e^{-t A}\right) = \delta_{m_1m_2}
\sum_{l=0}^{2L}\sum_{k=-[l/n]}^{[l/n]} \sum_{m_3=-L}^L  
\left( {C_{l,n k}}^{L,m_1;L,m_3}\right)^2 e^{-t\, l(l+1)}.
\ee
Now let me numerically analyze \eq{localtraces2znexp} for
the specific example of the fuzzy $S^2/Z_2$ with $L=10$. In Fig.\,\ref{fig55}, 
the behavior of the coefficient functions with $\nu=2$ and the insertion $\hat h_{0,0}$ is shown.
There clearly exists the stable region $0.06 \lesssim t \lesssim 0.2$. 
The values there may be well evaluated at $t=0.08$, and 
in Fig.\,\ref{fig6}, the $m$-dependence of $a_{h_{m,m},2j}^{3}(0.08)\ (j=0,1,2)$ is plotted. 
At small $|m|$, the values are stable under the change of $m$, and they take 
\bea
\label{locals2z2h0}
a^3_{h_{0,0},0}(0.08)&=&0.0238097, \cr
a^3_{h_{0,0},2}(0.08)&=&0.00793031, \\
a^3_{h_{0,0},4}(0.08)&=&0.00166333, \nonumber
\eea
for $m=0$. 

At small $|m|$, the operator $\hat h_{m,m}$ is well separated from the singularities on the poles 
in continuum theory, 
and it should pick out the smooth geometric contributions of the bulk.  
From the interpretation of \eq{identityhmm} explained above, each of the slices should have the same
weight, and \eq{locals2z2h0} should be compared with $1/(2L+1)$ of \eq{locals2z2h0c}:
\bea
\frac{1}{21} a_{0}({\rm bulk})&\approx & 0.0238095, \cr
\frac{1}{21} a_{2}({\rm bulk})&\approx & 0.00793651, \\
\frac{1}{21}a_{4}({\rm bulk})&\approx & 0.0015873. \nonumber
\eea
These continuum values are in good agreement with \eq{locals2z2h0}, and support the above
expectation.

\begin{figure}
\begin{center}
\includegraphics[scale=.65]{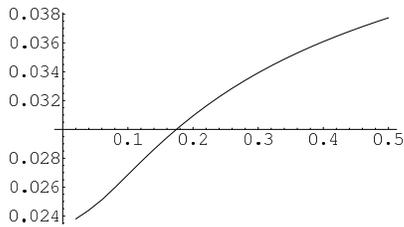}
\caption{The $t$-dependence of $a_{h_{10,10},0}^{3}(t)$. No stable regions can be found.}
\label{fig7}
\end{center}
\end{figure}

As shown in Fig.\,\ref{fig6}, the values depend largely on $m$ at large $|m|$, i.e.
near the singularities in continuum theory. 
This shows that the singularities in continuum theory are not concentrated in the $m=\pm L$ portion, but it
is distributed among some of them. The values fluctuate so largely that 
the distribution is not smooth and cannot be characterized.
In fact, the values for large $|m|$ are not reliable. This can be seen from that
the coefficient functions do not have the stable regions
for large $|m|$ as shown in Fig.\,\ref{fig7}.
This shows that 
the effective local geometry near the singularity in continuum theory cannot be well obtained 
through the method.

\begin{figure}
\begin{center}
\includegraphics[scale=.48]{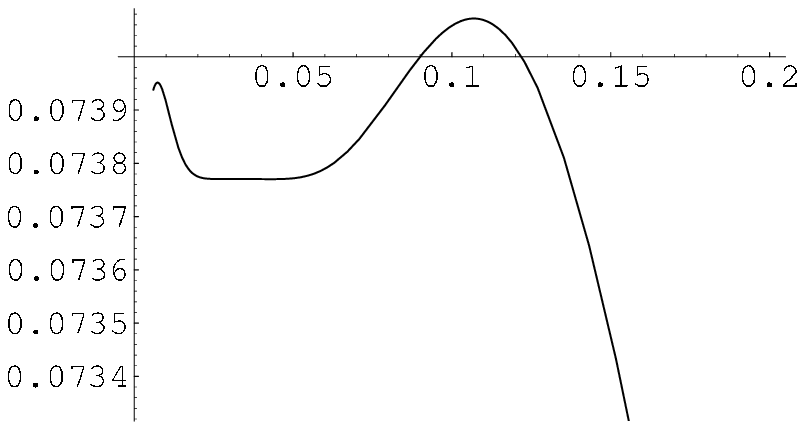} 
\includegraphics[scale=.48]{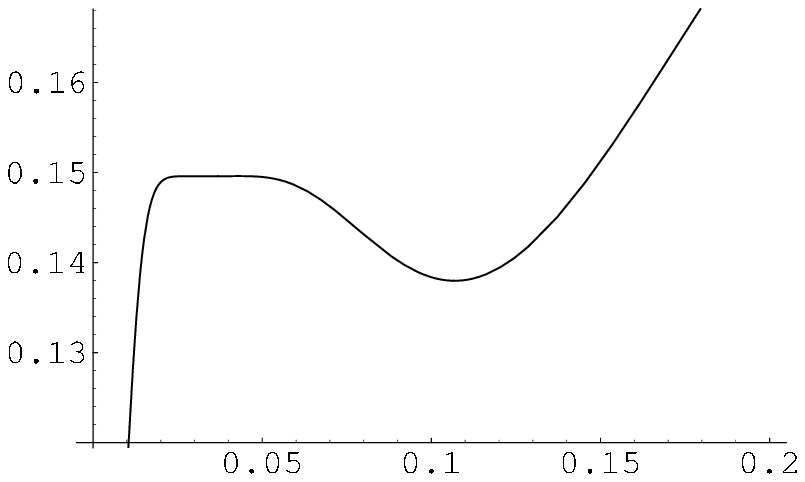}
\includegraphics[scale=.48]{locals2z2a2ms.eps}
\caption{The $t$-dependence of $a_{h_{22\leqslant},2j}^{4}(t)$ for the fuzzy $S^2/Z_2$ with $L=30$.} 
\label{fig8}
\end{center}
\end{figure}

On the contrary, it is possible to find the effective local geometric quantities associated with 
a certain broad area around the singularity in continuum theory. 
To see this, let me consider the insertion of the operator,
\be
\hat h_{m\leqslant}=\sum_{m_1=m}^L \hat h_{m_1,m_1},
\ee 
which contains the singularity in continuum theory.
In Fig.\,\ref{fig8}, the $t$-dependence of the coefficient functions is shown for the insertion
of $\hat h_{22\leqslant}$ and $L=30$. 
There clearly exists the stable region, and the effective geometric quantities can be found. 
These values may be well evaluated at $t=0.03$:
\bea
\label{locals2z2ms}
a^3_{h_{22\leqslant},0}(0.03)&=&0.0737705, \cr
a^3_{h_{22\leqslant},2}(0.03)&=&0.149593, \\
a^3_{h_{22\leqslant},4}(0.03)&=&0.0672558. \nonumber
\eea
From these values, subtracting the bulk contributions shown in \eq{locals2z2h0c}, 
the contributions associated with the singularity can be singled out:
\be
\label{s2z2asssing}
\begin{array}{rcl}
\Delta a^3_{h_{22\leqslant},0}(0.03)&=&0.0737705-\frac{9}{2 \cdot (2\cdot 30+1)}\approx 0.0000000 \cr
\Delta a^3_{h_{22\leqslant},2}(0.03)&=&0.149593-\frac{9}{6 \cdot (2\cdot 30+1)}\approx 0.125003, \\
\Delta a^3_{h_{22\leqslant},4}(0.03)&=&0.0672558-\frac{9}{30 \cdot (2\cdot 30+1)}\approx 0.0623378.
\end{array}
\ee
These are in good agreement with the analytical result \eq{locals2z2sing}.

\subsection{Fuzzy $S^1$ and $S^1/Z_2$}
\label{subsections1local}
The action \eq{actionh} in the limit $\alpha\rightarrow\infty$ chooses $\hat Y_{2L,m}\ (m=-2L,-2L+1,\cdots,2L)$ 
out of the scalar modes in the fuzzy $S^2$.  
These $\hat Y_{2L,m}\ (m=-2L,-2L+1,\cdots,2L)$ form the algebra of the fuzzy $S^1$.
It is expected that an appropriate linear combination of $\hat Y_{2L,m}$ can be used 
as the insertion operator to obtain the effective local geometric quantities. 
Especially, since $\hat Y_{2l,0}$ is the zero mode in the fuzzy $S^1$, its insertion 
must reproduce the global geometric quantities obtained in Section\,\ref{globals1}. 

\begin{figure}
\begin{center}
\includegraphics[scale=.65]{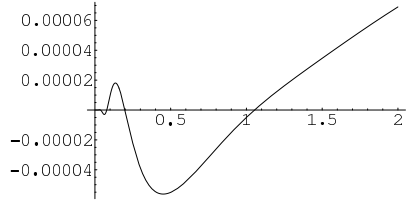} 
\caption{The $t$-dependence of $a_{Y_{2L,0},0}^{6}(t)$ for the fuzzy $S^1$ with $L=10$. No stable regions can
be found.} 
\label{fig9}
\end{center}
\end{figure}

Let me compute the coefficients in the formula \eq{htgen} for the insertion $\hat Y_{2l,0}$, 
\bea
{\rm Tr}_L \left( \hat Y_{2L,m}^\dagger \hat Y_{2L,0} \hat Y_{2L,m} \right)&=&\sum_{m_1,m_2,m_3} 
{C_{2L,0}}^{L,m_1;L,-m_2}(-1)^{L-m_2} {C_{2L,m}}^{L,m_1;L,m_3}{C_{2L,m}}^{L,m_2;L,m_3} \cr
&=&(4L+1) \left\{ \begin{array}{ccc} 2L&2L&2L \\ L&L&L \end{array} \right\} 
{C_{2L,m}}^{2L,0;2L,m},
\eea 
where $\{::: \}$ denotes the $6j$-Symbol, and I have used \eq{repY}, \eq{YhYcoeff} 
and some elementary properties of the Clebsch-Gordan coefficients \cite{Messiah, Varshalovich:1988ye}. 
Thus the heat trace is obtained as 
\be
{\rm Tr} \left( Y_{2L,0} \, e^{-tA}\right)
=(4L+1) \left\{ \begin{array}{ccc} 2L&2L&2L \\ L&L&L \end{array} \right\} 
\sum_{m=-2L}^{2L} {C_{2L,m}}^{2L,0;2L,m} e^{-m^2 t}.
\ee
I investigated the behavior of the coefficient functions for various $N$ and $L$, but could not find 
any stable regions as shown in Fig.\,\ref{fig9} for a specific case.
An apparent reason for this is that ${C_{2L,m}}^{2L,0;2L,m}$ does not behave like a constant as a function
of $m$, which is the case if the identity operator is inserted.
It can be also checked that the heat trace depends essentially on the insertion, which contradicts
the uniformity of $S^1$, as will be explained below.
Thus the fuzzy $S^1$ obtained by the reduction from the fuzzy $S^2$ discussed in \cite{Dolan:2003kq} does not
have the appropriate algebra for obtaining the local geometric quantities, 
although the global geometric quantities can be correctly obtained as in Section\,\ref{globals1}.
To produce the correct local geometric properties, one needs to allow the insertion operators $\hat h$ 
independent of $\hat Y_{2L,m}$ or to change the algebra itself. But it is not clear how to do.   

To circumvent the difficulty, let me consider a more direct approach. 
In continuum theory, a function on $S^1$ with length $2\pi$ 
can be expanded in a linear combination of $\phi_m\equiv e^{-imx}\ (m=0,\pm1,\cdots)$, 
where $x$ is the coordinate of the $S^1$. By introducing a cut-off parameter $L$,
the fuzzy $S^1$ can be defined by the algebra formed by $\phi_m\ (m=0,\pm1,\cdots,\pm L)$.
The simplest choice of the algebra would be obtained by projecting out the modes $\phi_m\ (|m|> L)$: 
\be
\label{algphi}
\hat \phi_m \hat \phi_n=\left\{ 
\begin{array}{cc} \hat \phi_{m+n},& -L\leq m+n \leq L \\ 0, & {\rm otherwise} \end{array}. 
\right.
\ee
This is a commutative {\sl non-associative} algebra. A non-associative algebra is usually not favored in physics, 
but plays the significant roles in constructing more than two-dimensional fuzzy spheres \cite{Ramgoolam:2001zx}.
The Laplacian in the fuzzy $S^1$ is defined by a linear operator $\Delta \hat\phi_m=-m^2 \hat \phi_m$. 
Now let me consider the insertion of a general element, $\hat h=\sum_{m=-L}^L c_m\hat\phi_m$, 
with the numerical coefficients $c_m$. Since
\be
\hat h\, \hat \phi_m=c_0\, \hat \phi_m+\cdots,
\ee
and \eq{htgen},
the heat trace with the insertion is obtained as 
\be
{\rm Tr}(h\, e^{-t A})= c_0 \sum_{m=-L}^L e^{-m^2 t}.
\ee
Thus the heat trace is proportional to the global one, irrespective of the insertion.
This is similar to the fuzzy $S^2$ as mentioned in Section \ref{localstwo},
and shows the uniformity of the fuzzy $S^1$. 

Now I consider a $Z_2$ reflection, which is defined by $U(\hat \phi_m)=\hat \phi_{-m}$. The fuzzy 
orbifold $S^1/Z_2$ is defined by taking the scalar modes invariant under the $Z_2$ transformation.
The algebra is formed by $\hat \phi_0,\, \hat\phi_m+\hat\phi_{-m}\ (m=1,\cdots,L)$. 
This should be the fuzzy analog of
the continuum $S^1/Z_2$ with the Neumann boundary condition at the boundaries.
From the expression of the $\delta$-function in continuum theory, 
a fuzzy analog of the $\delta$-function at $x=x_0\ (0\leq x_0 \leq \pi)$ can be defined as
\be
\hat \delta_L(x-x_0)\equiv \frac{1}{\pi}\hat\phi_0+\frac{1}{\pi} \sum_{m=1}^L \cos (mx_0) \, 
(\hat \phi_m+\hat \phi_{-m}).
\ee 
The heat trace with the insertion $h_{x_0}=\delta_L(x-x_0)$ is computed from \eq{htgen} and \eq{algphi} as
\be
{\rm Tr}\left( h_{x_0} e^{-t A}\right) =
\frac1\pi \sum_{m=0}^L e^{-m^2 t}+ \frac1\pi \sum_{m=1}^{[L/2]} \cos(2mx_0)e^{-m^2t}.
\ee

\begin{figure}
\begin{center}
\includegraphics[scale=.48]{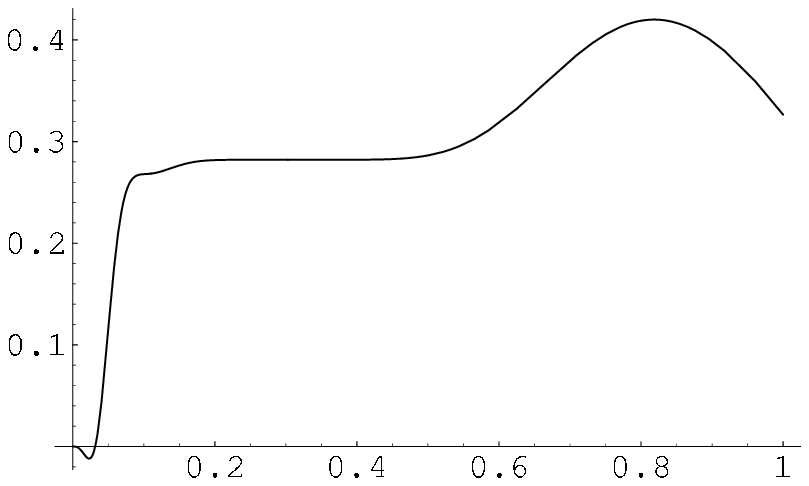} 
\includegraphics[scale=.48]{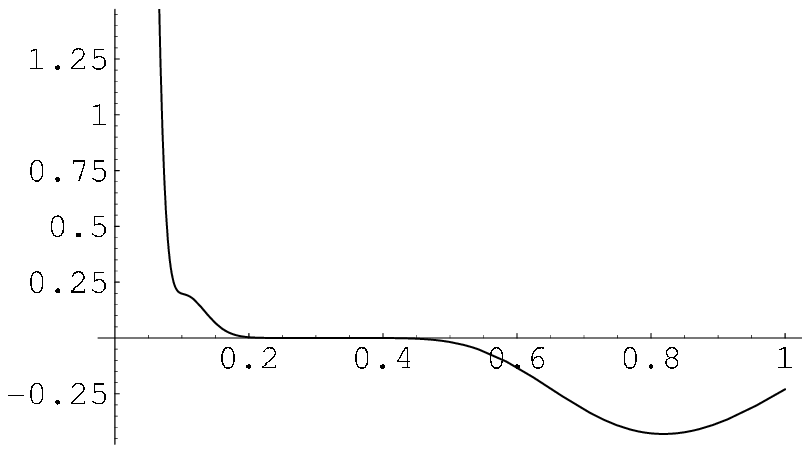} 
\includegraphics[scale=.48]{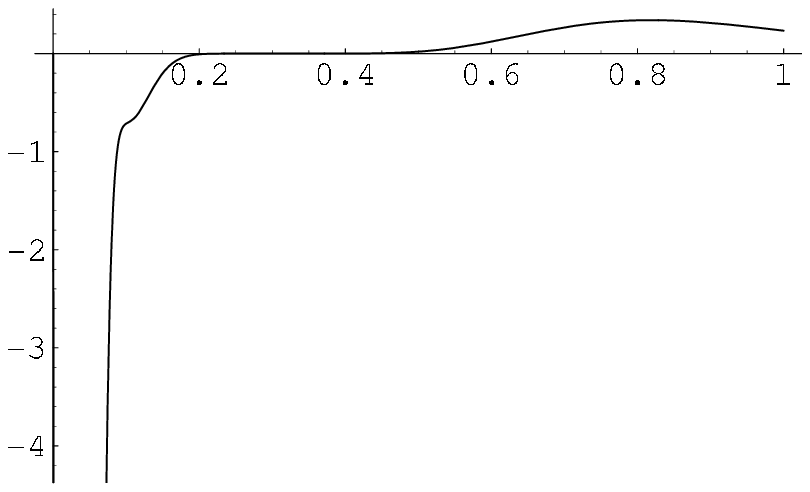} 
\caption{The $t$-dependence of $a_{h_{\pi/2},j}^{b,3}(t)\ (j=0,1,2)$ for the fuzzy $S^1/Z_2$ with $L=30$. 
There clearly exists the stable region $0.2 \lesssim t \lesssim 0.4$.} 
\label{fig10}
\end{center}
\end{figure}

\begin{figure}
\begin{center}
\includegraphics[scale=.48]{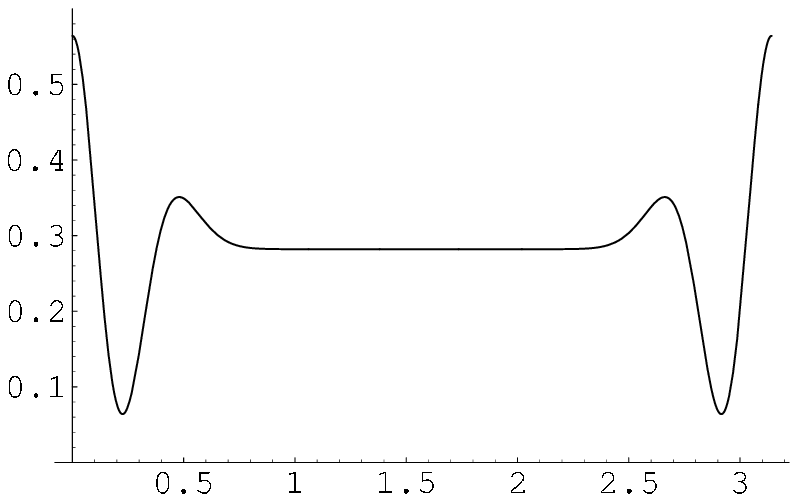} 
\includegraphics[scale=.48]{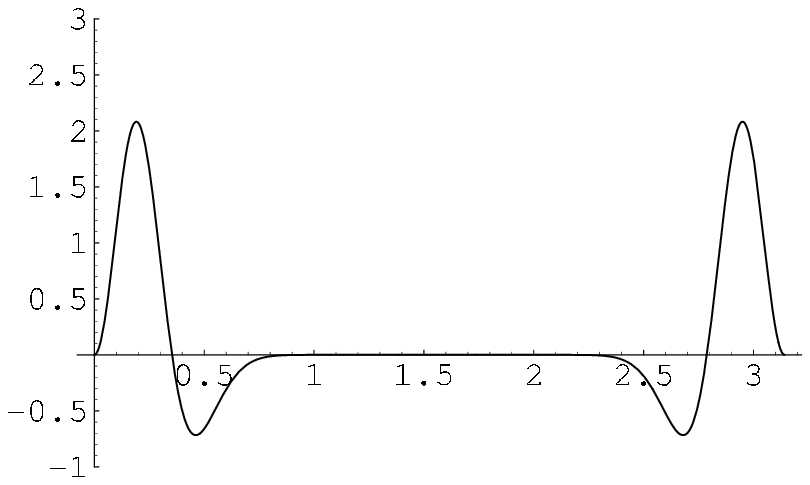} 
\includegraphics[scale=.48]{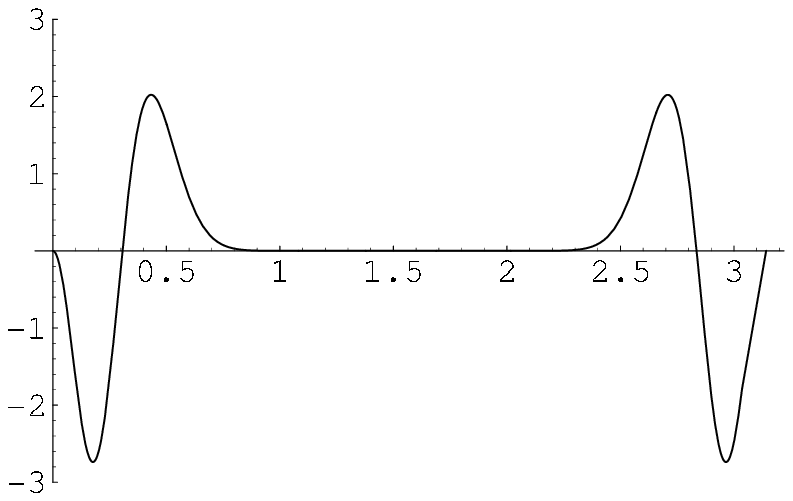} 
\caption{The $x_0$-dependence of $a_{h_{x_0},j}^{b,3}(0.25)\ (j=0,1,2)$ for the fuzzy $S^1/Z_2$ with $L=30$. 
The values fluctuate largely near $x_0=0,\pi$. } 
\label{fig11}
\end{center}
\end{figure}

\begin{figure}
\begin{center}
\includegraphics[scale=.65]{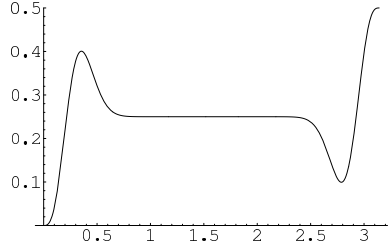} 
\caption{The $x_0$-dependence of $a_{h_{\leqslant x_0},1}^{b,3}(0.25)$ for the fuzzy $S^1/Z_2$ with $L=30$.} 
\label{fig12}
\end{center}
\end{figure}

Let me consider the specific example of the fuzzy $S^1/Z_2$ with $L=30$, and use 
the coefficient functions \eq{coefffunb} with $\nu=1$.
In Fig.\,\ref{fig10}, the $t$-dependence of the coefficient functions at the middle point $x_0=\pi/2$ 
of the $S^1/Z_2$ is shown. There clearly exists
the stable region $0.2\lesssim t \lesssim 0.4$.
The values there may be well evaluated at $t=0.25$, and 
the $x_0$-dependence of the coefficient functions at $t=0.25$ is shown in Fig.\,\ref{fig11}.
Like the fuzzy $S^2/Z_n$ in Section\,\ref{localstwo}, the coefficient functions of $S^1/Z_2$ 
fluctuate largely near the boundaries in continuum theory, 
and this suggests that the values are not reliable.
In fact, it can be checked that no stable regions can be found near 
the boundaries $x=0,\pi$. 
On the other hand, if the insertion operator is an integrated one including 
the boundary $x=0$, 
\be
\hat h_{\leqslant x_0}=\int_{0}^{x_0} dx_1 \ \hat h_{x_1}, 
\ee
there exists the stable region for $x_0$ sufficiently apart from the boundaries. In Fig.\,\ref{fig12}, 
the $x_0$-dependence of $a_{h_{\leqslant x_0},1}^{b,3}(0.25)$ is shown.  
As $x_0$ becomes larger, the value approaches $1/4$, stabilizes, and finally becomes $1/2$.
In continuum theory, this corresponds to the fact that $a_1$ gets the contribution $1/4$ 
from each boundary as in \eq{convals1oz2}.       

\subsection{A fuzzy line segment}
\label{localseg}
The action \eq{linesegaction} in the limit $\alpha\rightarrow\infty$ chooses the modes 
$\hat Y_{l,0}\ (l=0,1,\cdots,2L)$ from those in the fuzzy $S^2$. One can take another basis for these modes 
as $\hat h_{m,m}\ (m=-L,-L+1,\cdots,L)$. The analysis in \cite{Martin:2004dm, Sasakura:2004yr}
shows that the latter basis is more appropriate to give the insertion operators 
to analyze the local geometric properties. 
The parameter $m$ labels the `points' in the fuzzy line segment, and they
are placed in the order of $m$. The points $m=\pm L$ correspond to the two ends of the fuzzy line segment.   
Using \eq{htgen} and \eq{YhYcoeff}, the heat trace with the insertion $h_{m,m}$ is evaluated as
\bea
\label{localsegheatgen}
{\rm Tr}(h_{m,m} e^{-tA})&=&\sum_{l=0}^{2L} {\rm Tr}_L\left(\hat Y_{l,0}^\dagger \hat h_{m,m} \hat Y_{l,0}
\right)e^{-l(l+1)\,t} \cr
&=&\sum_{l=0}^{2L} \left({C_{l,0}}^{L,m;L,-m}\right)^2 e^{-l(l+1)\,t}.
\eea

\begin{figure}
\begin{center}
\includegraphics[scale=.48]{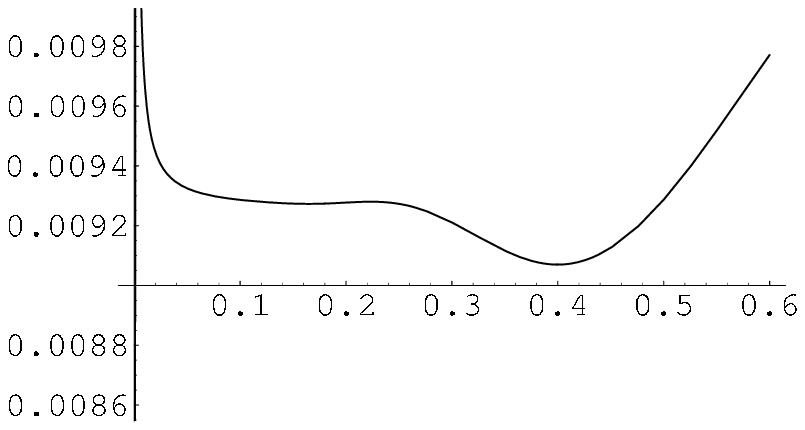} 
\includegraphics[scale=.48]{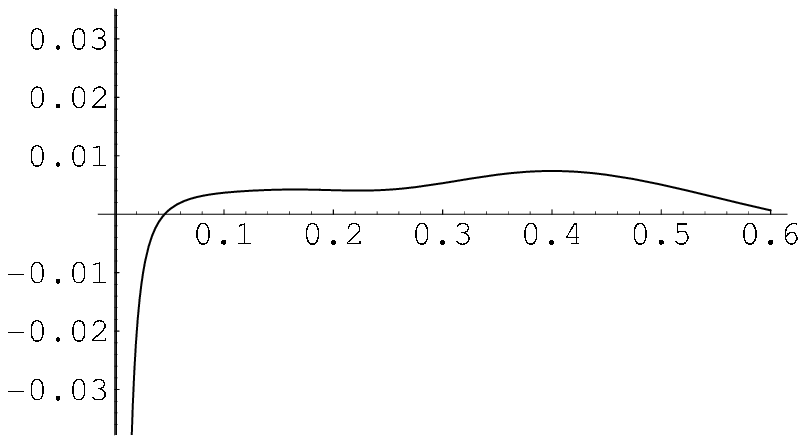} 
\includegraphics[scale=.48]{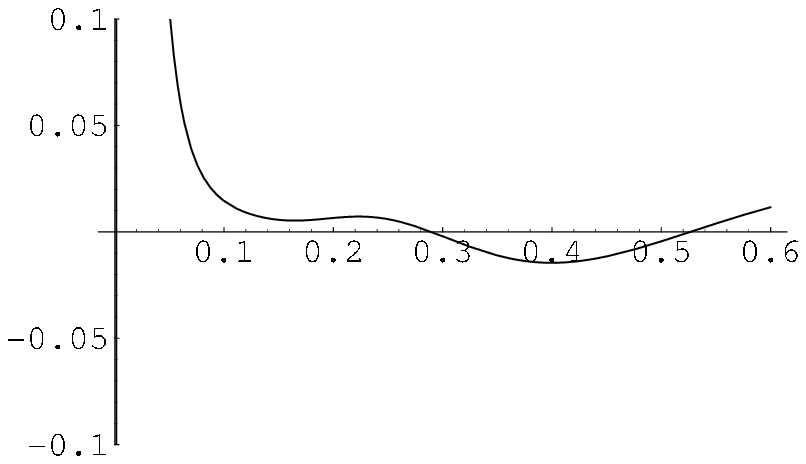} 
\caption{The $t$-dependence of $a_{h_{0,0},2j}^{6}(t)\ (j=0,1,2)$ for the fuzzy line segment with $L=30$. 
The stable region seems to exist around $t\sim 0.16$, but this is not clear.} 
\label{fig13}
\end{center}
\end{figure}

\begin{figure}
\begin{center}
\includegraphics[scale=.65]{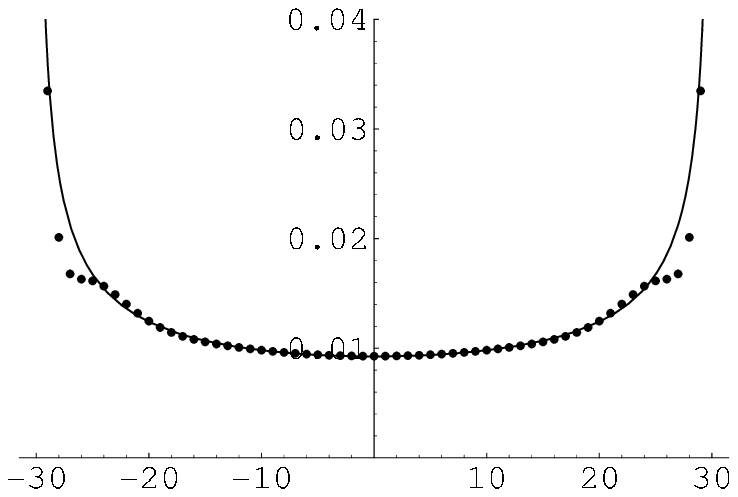} 
\hfil
\includegraphics[scale=.65]{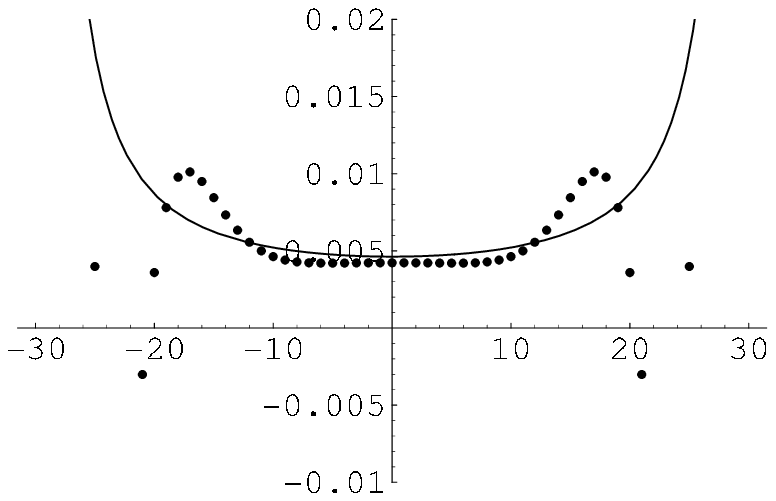} 
\caption{The $m$-dependence of $a_{h_{m,m},2j}^{6}(0.16)\ (j=0,1)$ for the fuzzy line segment with $L=30$
is shown with dots. The solid line is obtained from the continuum limit of the 
operator $A$.}
\label{fig14}
\end{center}
\end{figure}

Now let me numerically study the fuzzy line segment with $L=30$. 
The coefficient functions \eq{coefffun} are used, as in Section\,\ref{globalseg}.
In Fig.\,\ref{fig13}, the $t$-dependence
of the coefficient functions with $\nu=1$ and the insertion $h_{0,0}$ is shown. 
The existence of the stable region is not clear, but seems to exist around $t\sim 0.16$.    
In Fig.\,\ref{fig14}, the $m$-dependence of $a_{h_{m,m},2j}^{6}(0.16)\ (j=0,1)$ is shown, and is compared
with the continuum expression which will be given shortly. 
They agree well around the center, but the second coefficient $a_{h_{m,m},2}^{6}(0.16)$ disagrees largely 
apart from the center. It can be checked that when $|m|$ becomes larger, the existence of the stable region
becomes more obscure.

\begin{figure}
\begin{center}
\includegraphics[scale=.65]{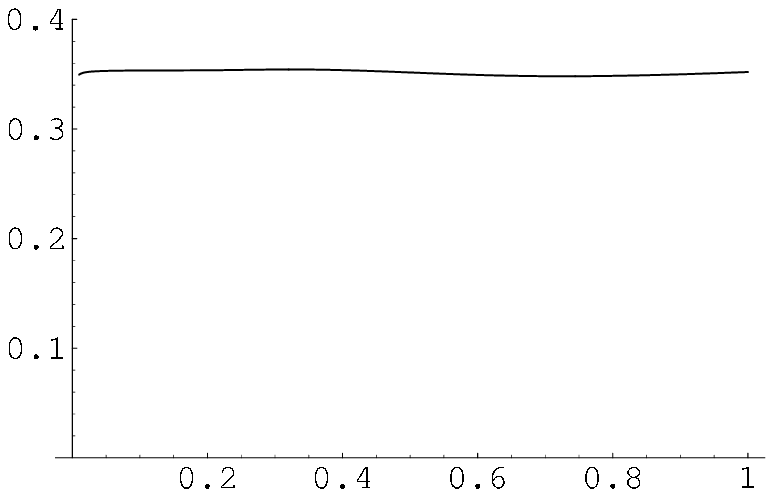} 
\hfil
\includegraphics[scale=.65]{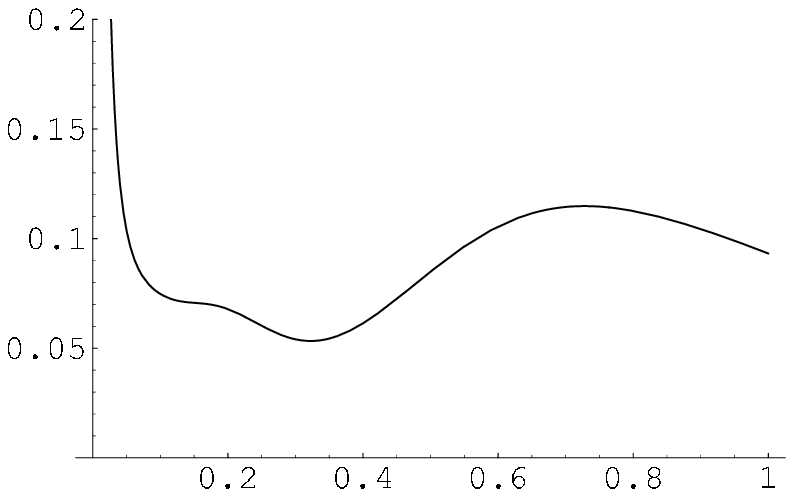} 
\caption{The $t$-dependence of $a_{h_{10 \leqslant },2j}^{6}(t)\ (j=0,1)$ for the fuzzy line segment with $L=30$.
The fluctuation of $a_{h_{10 \leqslant },2}^{6}(t)$ is large (right), while that of 
$a_{h_{10 \leqslant},0}^{6}(t)$ is within a few percent (left).
The stable region cannot be clearly found.}
\label{fig15}
\end{center}
\end{figure}

In the previous examples, 
the stable regions can be found when the insertion operator has the support
on a broad region around a singularity in continuum theory. Let me consider 
the insertion of $\hat h_{10 \leqslant }=\sum_{m=10}^{30} \hat h_{m,m}$.
The stable region cannot be clearly found, as shown in Fig.\,\ref{fig15}.
This suggests that the singularity on the boundary is so strong that its effect is not
localized near the boundary.

The above anomalous behavior of the numerical analysis can be supported by the following explicit evaluation
of the heat kernel trace.
Let me first consider the insertion $h_{0,0}$, which is localized about the center
of the fuzzy line segment. Substituting the explicit expressions of 
the Clebsch-Gordan coefficients \cite{Messiah, Varshalovich:1988ye}
into \eq{localsegheatgen}, the heat kernel trace with the insertion is given by
\bea
{\rm Tr}(h_{0,0} e^{-tA})  &=& \sum_{
 l=0,even}^{2L} \frac{(2l+1)(2L-l)!(l!)^2\left(\left(L+\frac{l}2\right)!\right)^2}{(2L+l+1)!
\left(\left(\frac{l}2\right)!\right)^4\left(\left(L-\frac{l}2\right)!\right)^2} e^{-l(l+1)t}.
\eea 
When $1\ll l \ll L$, the coefficients are approximated by
\be
\frac{(2l+1)(2L-l)!(l!)^2\left(\left(L+\frac{l}2\right)!\right)^2}{(2L+l+1)!
\left(\left(\frac{l}2\right)!\right)^4\left(\left(L-\frac{l}2\right)!\right)^2}
\sim \frac{2}{\pi L}.
\ee
Using this approximation and the Euler-Maclaurin Formula \eq{emformula}, 
the asymptotic behavior of the heat kernel trace is obtained as 
\be
\label{asymp00}
{\rm Tr}(h_{0,0} e^{-tA})  \sim \frac{1}{2\sqrt{\pi}L}t^{-1/2}+\cdots.
\ee
This asymptotic behavior is consistent with the effective dimension ($\nu=1$) assumed 
in the numerical analysis above, 
and the value $1/2\sqrt{\pi}L\approx 0.0094$ for $L=30$ is in good agreement with 
Fig.\,\ref{fig14}.
Let me next consider the insertion operator $h_{L,L}$, which is localized about the boundary. 
The explicit expression of the heat kernel trace with the insertion is given by
\be
{\rm Tr}(h_{L,L} e^{-tA})=\sum_{l=0}^{2L} \frac{(2l+1) \left( \left(2L\right)!\right)^2}{(2L+l+1)!(2L-l)!}
e^{-l(l+1)t}.
\ee  
When $1\ll l \ll L$, the coefficient is approximated by
\be
\frac{(2l+1) \left( \left(2L\right)!\right)^2}{(2L+l+1)!(2L-l)!}\sim \frac{2l+1}{2L}.
\ee
This $l$-dependence is what appears in the global analysis of the fuzzy two-sphere in Section\,\ref{globals2zn},
and changes the qualitative behavior of the asymptotic expansion from \eq{asymp00}:
\be
{\rm Tr}(h_{L,L} e^{-tA})  \sim \frac{1}{2L}t^{-1}+\cdots.
\ee
Therefore the effective dimension is two ($\nu=2$) on the boundary of the fuzzy line segment. 
Since the global analysis in Section\,\ref{globalseg} is consistent with $\nu=1$, this anomalous
property on the boundary must be canceled with that in the bulk. 
Therefore this anomalous property is not localized on the boundaries but must also 
exist in the bulk. This will invalidate the numerical analysis above based on $\nu=1$.
Thus the local analysis necessitates $\nu=2$, while $\nu=1$ is globally preferred, and 
they cannot be reconciled.

Next let us investigate the problem from the direct continuum limit of the operator $A$.
The continuum limit is discussed in \cite{Martin:2004dm, Sasakura:2004yr}, 
and is given by
\be
A=-\frac{d}{dx}x(1-x)\frac{d}{dx},
\ee
where $x$ is in the range $[0,1]$, and there are no constraints on the boundaries. 
This operator can be rewritten in the covariant form
\be
A=-(g^{xx} \nabla_x\nabla_x +E),
\ee
where $\nabla_x$ is the covariant derivative with $\Gamma_{xx}^x=(2x-1)/2x(1-x)$, and 
\be
\begin{array}{rcl}
\displaystyle g^{xx}&=&\displaystyle x(1-x), \\
\displaystyle E&=& \displaystyle \frac{1+4x-4x^2}{16x(1-x)}.
\end{array}
\ee
Thus $A$ contains the term $E$ additionally to the Laplacian. The $a_0$ is not changed by this additional term, 
but $a_2$ gets the additional bulk contribution \cite{Vassilevich:2003xt,Elizalde:1994gf, Gilkey:1995mj, Kirsten:2001wz}
\be
\frac{1}{(4\pi)^{\nu/2}} \int d^\nu x\,\sqrt{g}\, h\, E.
\ee  
Adding it to \eq{avalues}, the continuum expressions for $a_0(h),a_2(h)$ from the bulk 
are given by
\bea
a_0(h)_{bulk} &=& \frac{1}{(4\pi)^{1/2}} \int_0^1 dx\ \frac{h}{\sqrt{x(1-x)}}, \\
a_2(h)_{bulk}&=& \frac{1}{(4\pi)^{1/2}} \int_0^1 dx\ \frac{h}{\sqrt{x(1-x)}} \frac{(1+4x-4x^2)}{16 x(1-x) }.
\eea
Note that $a_0(1)_{bulk}$ is a well-defined quantity $\sqrt{\pi}/2$
and agrees with the analytical result \eq{globalsegasymp}, 
while the integration for $a_2(1)_{bulk}$ diverges at $x=0,1$ and is ill-defined. 
This divergence must be canceled with the boundary contributions in some way to reproduce 
the meaningful global result \eq{globalsegasymp}.
Therefore the boundary contributions cannot be localized on the boundaries, but must smear into 
the bulk to cancel the divergence.  
This provides another support to the anomalous behavior.
Since one may approximate
\be
h_{m,m}(x) \sim \frac{1}{2L+1}\delta\left(x-\frac{m+L}{2L}\right), 
\ee
the numerical analysis should be compared with the expressions,
\bea
a_0(m)_{cont.}&=&\frac{1}{(2L+1)\sqrt{4\pi x(1-x)}}, \\
a_2(m)_{cont.}&=&\frac{1+4x-4x^2}{16(2L+1)\sqrt{4\pi} x^{3/2}(1-x)^{3/2}},
\eea
where $x=(m+L)/(2L)$. These are the solid lines in Fig.\,\ref{fig14}.

\section{Summary and discussions}
\label{summary}
The idea that generalization of space can be obtained in terms of algebra is attractive 
\cite{Connes,Madore:aq,Landi:1997sh}. 
Then the geometry in such a generalized space is a secondary product encoded in the algebra. 
In the analogy of the classical particle mechanics, 
the effective geometry may be defined through the low-frequency 
dynamics of the fields in such a space. This is in accordance with the spirit of \cite{Chamseddine:1996zu}.   
  
In the previous paper \cite{Sasakura:2004dq}, I discussed a method of obtaining the global geometric
quantities of compact fuzzy spaces from the approximate power-law expansion of the heat kernel trace. 
In this paper I applied the method to the heat kernel trace with the insertion of local operators to check 
whether the effective local geometric quantities can be obtained through the method. 
In all the simple fuzzy spaces studied in this paper except the fuzzy line segment, 
the effective local geometric quantities obtained through the method are reasonable and support its validity.

The method does not provide any effective local geometric quantities near a singularity in continuum theory 
in a well-defined way, and provides them only for a certain broad range containing it. 
The physical interpretation of this fact would be that the effective geometry integrated 
over a certain broad range around a singularity in continuum theory is the only observable, 
while the effective geometry itself is not observable near it.
The fuzzy line segment gives an interesting counter example for the applicability of the 
present method. The metric singularity on the boundaries of the line segment in continuum theory 
is so strong that the method cannot give the local geometric quantities even well apart 
from the boundaries.

As for the global quantities, the heat kernel trace is fully determined by the spectra of the Laplacian,
while the algebraic relations among the modes are additionally needed to obtain the local geometric
quantities. The formulation of the fuzzy $S^1$ by the reduction from the fuzzy $S^2$ 
gives the spectra to produce the correct global geometric quantities 
\cite{Sasakura:2004dq}. However, the algebra of the modes is inappropriate to 
produce the local geometric quantities. 
On the other hand, a more direct approach with a non-associative 
algebra, which is also used in formulating more than two-dimensional fuzzy spheres \cite{Ramgoolam:2001zx}, 
produces the correct local geometric quantities. Since fuzzy spaces with non-associative algebra 
contain various physically interesting spaces,  
it would be interesting to apply the present method to study their effective geometry. This would 
be helpful in understanding the gravitational aspects and 
formulating the evolution of fuzzy spaces \cite{Sasakura:2004yr,Sasakura:2004vm,Sasakura:2003ke}.

In the present method, the local geometric quantities of fuzzy spaces depend both on the Laplacian and 
the algebra of the modes. In Section\,\ref{subsections1local}, the fuzzy $S^1$ is 
defined by the eigenvalues of the Laplacian and the 
non-associative algebra of the eigenmodes.
No intrinsic relations between the Laplacian and the algebra are assumed there.
However, since the algebra of the modes describes a certain aspect of the structure of a fuzzy space, it would 
be more reasonable to determine the Laplacian from the algebra through a principle.
One way to achieve this would be starting with the non-commutative differential geometry 
\cite{Connes,Madore:aq,Landi:1997sh}.
Another unsatisfactory treatment in this paper is that the choices of the local insertion operator $h$ 
are not fully derived. For the simple fuzzy spaces studied in this paper, the natural intuitive choices work 
well, and there would be no many other reasonable choices. But this will not be true for general fuzzy spaces. 
It is more desirable to be able to construct the local insertion operators starting from the algebra of a fuzzy
space without any speculations.  

In this and the previous \cite{Sasakura:2004dq} papers, the method is applied to 
extract the effective geometric quantities in compact fuzzy spaces. 
The fundamental physical assumption underlying the method is that there exists an effective field theory 
described in a usual manner in low-frequency. 
This will hold in more general regularized spaces such as $q$-deformed spaces, lattice theories and so on.
These spaces may also have statistical fluctuation, which may be interpreted as a quantum gravity effect.
The heat kernel expansion is known to have various applications \cite{Vassilevich:2003xt,Elizalde:1994gf, Gilkey:1995mj, Kirsten:2001wz}. 
Thus the present method would have various applications in various regularized spaces. 

\vspace{.5cm}
\noindent
{\large\bf Acknowledgments}\\[.2cm]
The author was supported by the Grant-in-Aid for Scientific Research No.13135213 and No.16540244
from the Ministry of Education, Science, Sports and Culture of Japan.

\end{document}